\documentclass{aa}  
\usepackage{amssymb}
\usepackage{amsfonts}
\usepackage{amsbsy}
\usepackage{amsmath}
\usepackage{pdflscape}
\usepackage{graphicx}
\usepackage{txfonts}
\usepackage{natbib,twoopt}
\usepackage[breaklinks=true]{hyperref} %% to avoid \citeads line fills

\newcommand{\CI}{C{\sc I}}
\newcommand{\CII}{C{\sc II}}
\newcommand{\HII}{H{\sc II}}
\newcommand{\HI}{H{\sc I}}

\bibpunct{(}{)}{;}{a}{}{,}             %% natbib format for A&A and ApJ
\makeatletter
  \newcommandtwoopt{\citeads}[3][][]{\href{http://adsabs.harvard.edu/abs/#3}%
   {\def\hyper@linkstart##1##2{}%
     \let\hyper@linkend\@empty\citealp[#1][#2]{#3}}}
  \newcommandtwoopt{\citepads}[3][][]{\href{http://adsabs.harvard.edu/abs/#3}%
    {\def\hyper@linkstart##1##2{}%
     \let\hyper@linkend\@empty\citep[#1][#2]{#3}}}
  \newcommandtwoopt{\citetads}[3][][]{\href{http://adsabs.harvard.edu/abs/#3}%
    {\def\hyper@linkstart##1##2{}%
     \let\hyper@linkend\@empty\citet[#1][#2]{#3}}}
  \newcommandtwoopt{\citeyearads}[3][][]%
    {\href{http://adsabs.harvard.edu/abs/#3}
    {\def\hyper@linkstart##1##2{}%
     \let\hyper@linkend\@empty\citeyear[#1][#2]{#3}}}
\makeatother

\begin{document}

   \title{Extreme conditions in the molecular gas of lensed star-forming galaxies at z$\sim$3}
 %  \subtitle{\large Detection of Atomic Carbon and CO(7-6) lines}

   \author{Paola Andreani\inst{1}
          \and
          Edwin Retana-Montenegro\inst{2}
         \and
          Zhi-Yu Zhang\inst{1,3}
         \and
          Padelis Papadopoulos\inst{4,5,6}
         \and
          Chentao Yang\inst{7}
         \and
          Simona Vegetti\inst{8}
          }

   \institute{European Southern Observatory, Karl-Schwarzschild-Stra\ss e 2, 85748 Garching, Germany
\email{pandrean@eso.org} 
         \and
             Leiden Observatory, Leiden University, P.O. Box 9513, 2300 RA, Leiden,The Netherlands
\email{eretana@strw.leidenuniv.nl}
         \and
         Institute of Astronomy, University of Edinburgh, Royal Observatory, Blackford Hill, Edimburgh EH9 3HJ, UK
        \and
         Department of Physics, Section of Astrophysics, Astronomy and Mechanics, Aristotle University of Thessaloniki, Thessaloniki, Macedonia, 54124, Greece
        \and
        Research Center for Astronomy, Academy of Athens, Soranou Efesiou 4, GR-115 27 Athens, Greece
        \and
        School of Physics and Astronomy, Cardiff University, Queen’s Buildings, The Parade, Cardiff, CF24 3AA, UK
         \and
         European Southern Observatory, Alonso de Cordova, 3107, Vitacura, Casilla 19001, Santiago de Chile, Chile
         \and
         Max-Planck Institut f\"ur Astrophysik, Boltzmann strasse , 85748 Garching, Germany
 %            \thanks{}
             }

   \date{Received ; accepted }

% \abstract{}{}{}{}{} 
% 5 {} token are mandatory
 
  \abstract
  % context heading (optional)
  % {} leave it empty if necessary  
%   {Line spectroscopy is the fundamental tool to investigate the physical conditions of the gas in the interstellar medium (ISM).}
  % aims heading (mandatory)
{}
   {Atomic carbon can be an efficient  tracer of the molecular gas mass, and when combined to the detection of high-J and low-J CO lines it yields also a sensitive probe of the power sources in the molecular gas of high-redshift galaxies.}
%{}
  % methods heading (mandatory)
   {The recently installed SEPIA 5 receiver at the focus of the APEX telescope has opened up a new window at frequencies 159 -- 211 GHz allowing the exploration of the atomic carbon in high-z galaxies, at previously inaccessible frequencies from the ground.
We have targeted three gravitationally lensed galaxies at redshift of about 3 and conducted a comparative study of the observed high-J CO/\CI ~ratios with well-studied nearby galaxies.}
  % results heading (mandatory)
   {Atomic carbon (\CI(2--1)) was detected in one of the three targets and marginally in a second, while in all three targets the $J=7\to6$ CO line is detected.}
   {The  CO(7--6)/\CI(2--1),  CO(7--6)/CO(1--0) line ratios and the CO(7--6)/(far-IR continuum) luminosity ratio are compared to those of nearby objects. A large excitation status in the ISM of these high-z objects is seen, unless differential lensing unevenly  boosts the CO line fluxes from the warm and dense gas more than the CO(1--0), \CI(2--1), tracing a more widely distributed cold gas phase. 
We provide estimates of total molecular gas masses derived from the atomic carbon and the carbon monoxide CO(1--0), which within the
uncertainties turn out to be equal.}
  % conclusions heading (optional), leave it empty if necessary 
 %  {}

   \keywords{galaxies: starburst – galaxies: ISM –  ISM: molecules – ISM: abundances – submillimeter: galaxies  –  techniques: spectroscopic
               }
\titlerunning{Atomic carbon in high-z galaxies}
\authorrunning{P. Andreani et al.}
\maketitle

%
%-------------------------------------------------------------------

\section{Introduction}

The initial conditions of  star formation inside molecular clouds are set by the gas temperature,  density, dynamical state,  and the local radiation field because they determine the free-fall time and the Jeans mass. Understanding these conditions is essential to fully understand the physics that drives the bulk of the star formation, especially during the early phase of the Universe.

Atomic carbon (\CI) forbidden fine-structure lines\footnote{The CI excited fine levels $^3$P$_1$ and $^3$P$_2$ lie 23.6 and 62.4 K above the ground state ($^3$P$_0$)
 and are therefore easily populated by particle collisions in the cold interstellar medium (ISM). Two magnetic-dipole transitions are allowed between the fine-structure levels:
$^3$P$_1$ $\to$ $^3$P$_0$ (\CI(1--0)) has a rest frequency of 492.1607 GHz, while $^3$P$_2$ $\to$ $^3$P$_1$ (\CI(2--1)) of 809.3435 GHz.}
have been proven to be a good tracer of  total
molecular gas (H$_2$),  even  better than  rotational transitions of carbon monoxide (CO) employed by  traditional studies  under a wide range of physical conditions \citep{ger+00,pap+04b,tom+14}. They have  emergent flux densities per H$_2$ column density higher than those of the low-CO rotational lines even at low metallicity and they are fully concomitant with CO line emission \citep{ger+00,pap+04b,ala+13}. \par\noindent

There is a a positive K-correction of the two \CI~lines  versus the low-J  CO transitions,  yet  their  similar excitation  characteristics
 start  giving an  advantage to  the former  at z$\ga  0.5$. \CI~lines can  remain well-excited for cooler,  lower density molecular
gas. This allows the use of \CI~lines as tracers of the molecular gas mass and gas-rich galaxy  dynamics over  a  much larger  fraction  of cosmic  look-back time.  Finally, they   are  optically  thin  (i.e.  there is no   need  for  an
X$_{CO}$-like factor), with only the $\rm [C/H_2]$ abundance as their main source of uncertainty (an  uncertainty common to all species that
are not H$_2$, e.g. $^{13}$CO lines or dust emission as $\rm H_2$ gas mass estimators).

\par\noindent

The  two high-J  CO lines  with frequencies  similar to  the \CI~lines (J=4--3, 7--6), are typically bright  in star-forming  galaxies, but
only trace   the dense  and warm H$_2$  gas near  star-forming regions, while the low-J  CO and \CI ~lines remain bright  in relatively diffuse ($10^3$ cm$^{-3}$) and colder ($\sim$ 10--30 {\rm k}) gas phases.  Furthermore, it is known  that C is not limited only to a  thin  C$\to$ CO  transition  zone  of far-UV (FUV) illuminated  clouds,  as
predicted  by photon  dissociation region  (PDR) models,  but remains concomitant  with  the  entire  CO-rich cloud, and its  distribution  may
actually      even      go      beyond     the      CO-rich      parts \citep[e.g.][]{pap+04b,bis+15}.  Thus,  CO(high-J)/\CI  ~line  luminosity
ratios  can  be  considered    good  proxies  of the  (warm,  dense,  star-forming gas)/(total  H$_2$ gas)  mass fraction,  even as  degeneracies
between temperature,  density, and line optical depths prevent  us from attributing actual mass fractions to these line ratios.
     
\par\noindent
Finally, although  the emission of singly ionised carbon fine structure transition, \CII, is much brighter than \CI, and has already been detected in high-z galaxies \citep[z$\sim$5--7; e.g.][]{deb+14,knu+16,hay+17,bra+17},  its emission traces ionised gas (\HII) and atomic gas (\HI) dominated regions, making the interpretation as $\rm H_2$ gas mass distribution tracer cumbersome.

At very high redshifts CI lines are also much less affected by the cosmic microwave background than the low-J CO lines \citep{zha+16}. Moreover, the \CI~abundance is less sensitive to the astrochemical effects of enhanced cosmic ray (CR) densities expected in starbursts (e.g. submillimetre galaxies (SMGs)) \citep{bis+15}.
CRs can effectively destroy CO throughout H$_2$ clouds, leaving C (but not much \CII), and unlike FUV photons that only act on the surface of the H$_2$ clouds and produce \CII, CRs destroy CO volumetrically and can make the clouds CO-invisible \citep{bis+17}.

\noindent
At redshifts z$\sim$2--4, the \CI ~lines are redshifted into observable windows from the ground. 
Several galaxies have been observed
\citep[][and references therein]{wei+05,pet+04,wal+11,ala+13,omo+13,yan+17,bot+17,pop+17},
showing that the SMGs share properties with local ultraluminous infrared galaxies and with less compact, local
starburst galaxies, providing new evidence that many SMGs have extended star formation distributions.
The total molecular mass inferred from these \CI ~observations was found to be in disagreement with that obtained via
the traditional conversion factor between CO mass and H$_2$ \citep[e.g.][]{pop+17}.

Here we report the observations of \CI(2--1) and the CO(7--6) lines towards three ${\rm z\sim}$3 lensed SMGs redshifted into the side bands of the ALMA Band 5 receiver  installed at the APEX telescope, and offered to the science community since 2015.
These targets have been selected from the HATLAS catalogue of bright submillimetre lensed objects \citep{bus+13} because of their brightness and their spectroscopic redshift, which allows the observations of the \CI(2--1) within the SEPIA 5 receiver bandwidth.

\section{SEPIA5/APEX observations}

The observations were carried out with the SEPIA Band 5 receiver \citep{bel+17} at the Atacama Pathfinder EXperiment  \citep[APEX; ][]{APEX}. %\citetads{2006A&A...454L..13G}.
The receiver covers the frequency range 159 -- 211 GHz.  The lower and upper sideband (LSB and USB) are separated by 12 GHz, and each sideband is recorded by two  eXtended bandwidth Fast Fourier Transform Spectrometer (XFFTS) units of 2.5 GHz each, with a 1GHz overlap.

The spectra have been smoothed to a spectral resolution of 20 km$s^{-1}$ and the beam size is 35$^{\prime\prime}$. We use a Jy/K factor of 34 to convert between antenna
temperature T$^*_{\rm A}$ and flux density assuming point sources \footnote{http://www.eso.org/sci/activities/apexsv/sepia/sepia-band-5.html} \citep{bil+12,imm+16}.

We conducted observations towards G12v2.43 during science verification (Project ID: 095.F-9803; PI: Andreani) and then for all the three sources during normal PI observing time  using  the  wobbler  switch  mode  with  precipitable water vapour (pwv) between 0.4 and 1.8mm. The observations were carried out for a total time of 9.6 hours, with 1, 2, 1.5 hours on source. Pointings were regularly checked on RW-LMi and IRC+10216, and a calibration scan was taken every 10 min. Data were reduced using CLASS/Gildas2. The SEPIA receiver was tuned at 189.3906, 195.42896, and 196.2085 GHz in the LSB, respectively. This corresponds to the
redshifted frequencies between the two restframe frequencies 809.3 and 806.65 GHz of the \CI(2--1) and CO(7--6) lines.

We removed a linear baseline from each individual spectrum before averaging the data. The rms reached are 1.0, 0.33, 0.7 mK, respectively, at a 50 km s$^{-1}$
resolution.

%-------------------------------------- Two column figure (place early!)\begin{landscape}

\begin{table*}
\caption{Observed line fluxes, velocities, and luminosities}             
\label{table:data}      
%\centering
\hskip -0.7cm          
\begin{tabular}{c c c c c c c c c c c}     % 6 columns 
\hline\hline       
\\
Galaxy & redshift & CO(7--6) & $\Delta v$ & \CI(2--1) & $\Delta v$ & CO(1--0) \tablefootmark{a,b} & $\Delta v$\tablefootmark{a,b}  &
$L_{\rm FIR}$\tablefootmark{c} & {$\rm \mu$} \tablefootmark{c}\\ 
            & & $(Jy~km/s$) & ($km/s$) & ($Jy~km/s$) &  ($km/s$) & ($Jy~km/s$) & ($km/s$) & 
( $10^{12} L_\odot$) &  \\

\hline    
\\
 J114637.9-001132  & 3.2588& 36.1$\pm$5.3 & 833$\pm$295 & 14.8$\pm$3.6 & 300$\pm$100 & 0.99$\pm$0.16 & 680$\pm$80  & 
14.2$\pm$1.2 & $9.5\pm0.6$ \\ 
(G12v2.30) & & & & &  & & & &\\\
 J113526.3-014605 & 3.1276 & 7.1$\pm$1.4 & 145$\pm$37 & $<$6 & $<$300 &  0.35$\pm$0.08 & 210$\pm$30 &
 7.5$\pm$4.8 & $17\pm11$ \\
 (G12v2.43) & & & & &  & & & &\\\
 J133800.8+245900  & 3.1112 & 32.6$\pm$6.6 & 845$\pm$202 & 8.2$\pm$5.8 & 542$\pm$507 &   3.30$\pm$ 0.50&  560$\pm$70 &
 12.3$\pm$1.4 & $13\pm7$ \\ 
(NBv1.78) & & & & &  & & & &
\\
\hline           
    
\end{tabular}
\tablefoot{\\
\tablefoottext{a}{CO(1--0) data from \citet{har+12}}
\tablefoottext{b}{for NBv1.78 data are relative to the CO(3--2) transition, from \citet{omo+13}}
\tablefoottext{c}{de-magnified values and magnification values from \citet{zha+18}}
}
\end{table*}

   \begin{figure}[t]
  \centering
   \resizebox{\hsize}{!}{\includegraphics[angle=90]{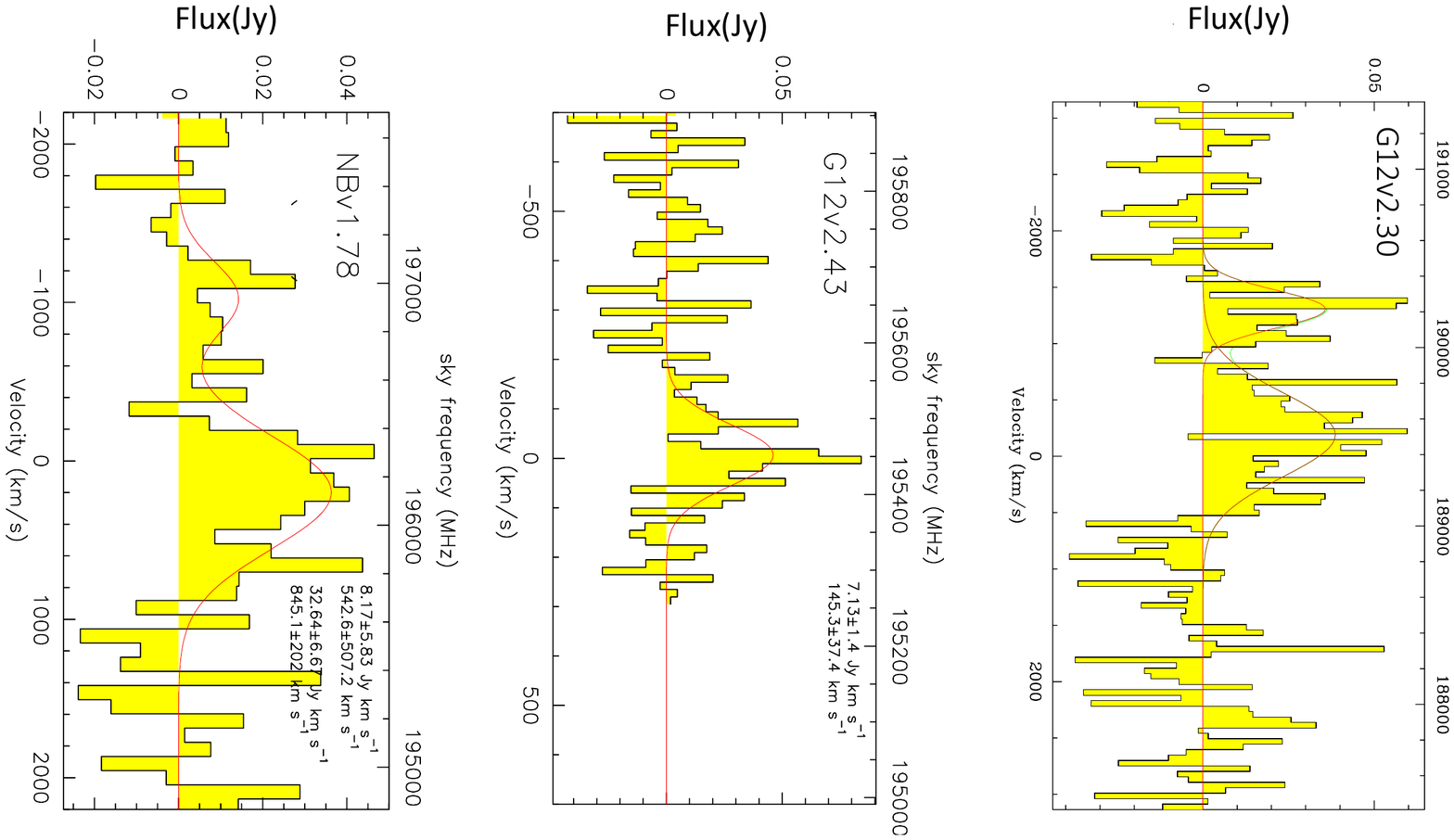}}  % \includegraphics[bb=0 0 900 900,angle=90,width=\hsize,clip]{Figure1.pdf}
     \caption{Spatially integrated APEX/SEPIA5 spectra of the three lensed sources, targets of this work. The red lines represent the Gaussian fitting to the
emission lines (CO(7--6) at $\sim$ zero velocity and \CI(2--1)). Zero velocity is set to the CO(1--0) and CO(3--2) line sky frequency according to the previously measured
spectroscopy redshifts given in Table~\ref{table:data} }\label{Spectra} 
%\centering\protect\caption{\label{Spectra} }
 \end{figure}

%--------------------------------------------------------------------
\section{Results}

\subsection{Line ratio diagnostics}

The spectra extracted from the SEPIA5 receiver are shown in Figure~\ref{Spectra}. The CO(7--6) line is clearly detected in all three sources 
and the corresponding fluxes and line widths resulting from the Gaussian best fit are listed in Table~\ref{table:data}.

The \CI(2--1) line is detected at a 4$\sigma$ level in G12v2.30, marginally ($\sim 2\sigma$) in NBv1.78 and not detected in G12v2.43. The corresponding fluxes or upper limits and line widths are listed in
Table~\ref{table:data}, together with the CO(1--0) fluxes taken from \citet{har+12}, CO(3--2) in NBv1.78 flux from \citet{omo+13}, the IR luminosities and magnification values from \citet{zha+18}. 

We use the line luminosity ratios ${\rm \frac{L^{\prime} (CO(7-6))}{L^{\prime} (CO(1-0))}}$ and ${\rm \frac{L^{\prime} (CO(7-6))}{L^{\prime} (\CI(2-1))}}$  as proxies for the ratio of the warm (T$\sim 100-150$K), dense (n$>10^4 cm^{-3}$) gas mass, $\rm M_{WD}$, to the total H$_2$ gas, $\rm M_{tot}(H_2)$,  $\rm M_{WD}/M_{tot}(H_2)$ mass fractions. Because the line radiative transfer model degeneracies prevent  the computation of actual mass fractions, we use other galaxy-averaged
ratios to place the  gas conditions in the submm galaxies in perspective, and even examine whether the observed line ratios are compatible with only FUV photon-driven energy sources for their ISM. This is important  since PDRs 
have often been found inadequate to account for the global CO line excitation measured in many starbursts  \citep[][and references therein]{pap+12a}. In placing our observed line ratios among the same ones measured for other galaxies where much more molecular and atomic line data allow a determination of their  $\rm M_{WD}/M_{tot}(H_2)$ mass fractions (and whether these are sustainable
by FUV radiation fields), we can circumvent to some degree the above-mentioned radiative transfer modelling uncertainties
(which we do not attempt here), and examine the type of prevailing ISM power source in our galaxies as well.

\begin{figure}[h]
   \centering{}
   \resizebox{\hsize}{!}{\includegraphics{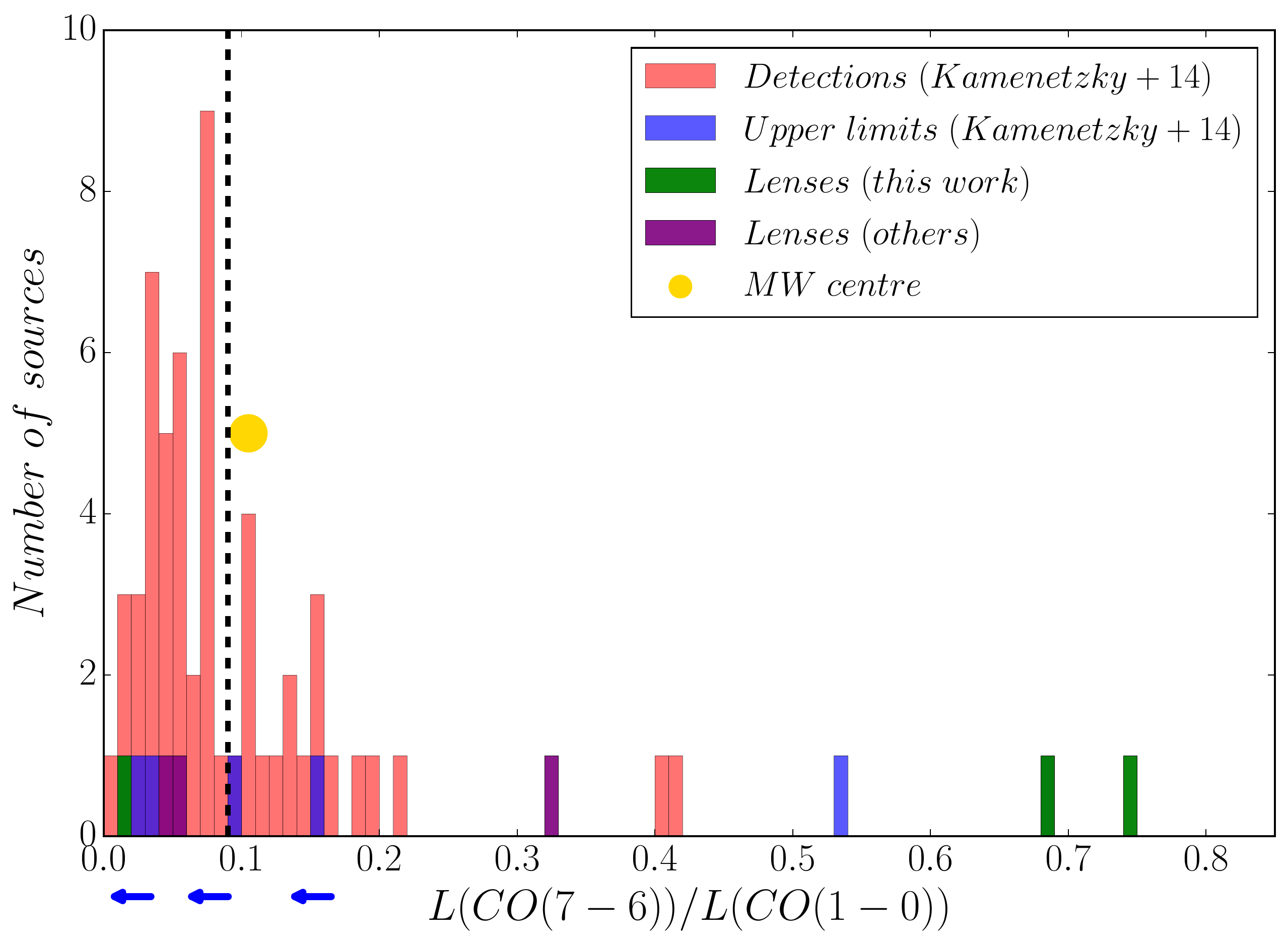}}
   \caption{Number of sources with a given CO(7--6)/CO(1--0) luminosity ratio (in units of K km~s$^{-1}$ pc$^2$). Data from this work are marked in green; red  and blue  bars refer to detections and upper limits, respectively, taken from \citet{kam+14}. The value shown here for NBv1.78 is CO(7--6)/CO(3--2) ($\sim 0.8$) and the corresponding CO(7--6)/CO(1--0) is expected to be lower by a factor of $\sim$3. Data from other similar lensed targets are taken from \citet{ote+17,yan+17} (shown in purple). The 
 Milky Way value is from \citet{fix+99} and corresponds to the centre of the Galaxy; values for the inner and outer Galaxy are upper limits ($<0.04$ and $<0.07$, respectively) and are not shown here. Values between 0.15 and 0.4 correspond to LIRGs ($\rm L(IR)>11.5 L_\odot$), values between 0.05 and 0.15 correspond to IRGs with ($\rm 11.0 <L(IR)<11.5 L_\odot$) and submillimetre galaxies, while lower values to galaxies with ($\rm L(IR)<11.0 L_\odot$). 
Lenses detected in this work show values higher than those of the local ULIRGs. The black dashed vertical line marks the values below which the objects are FUV photon dominated from those above which are non-FUV photon heated (like NGC253, NGC6240, Arp220, the Galactic Centre, etc.). }
              \label{CO76-CO10}%
    \end{figure}

In Figure~\ref{CO76-CO10} we compare the ratio $\rm \frac{CO(7-6)}{CO(1-0)}$ between the line luminosities in  ${\rm K~km~s^{-1} pc^2}$ for the sources listed in Table~\ref{table:data} with the same values for nearby galaxies \citep{kam+14,ros+15,lu+17,kam+17}. For the extended objects we consider those whose fluxes are aperture-corrected \citep{kam+14}. The CO line luminosity has been computed following \citet{sol+97}.
We add in Figure~\ref{CO76-CO10} the corresponding values of the Milky Way taken from the COBE observations for the Galactic Centre (GC) \citep{fix+99}. The corresponding values of similar lensed sources taken from the literature \citep{ote+17,yan+17,wal+11} are also shown for comparison. 

The histogram in Figure~\ref{CO76-CO10} can be used to examine the dominant power input of molecular gas reservoirs,  whether it is provided by  FUV photons  or  dominated by some other mechanisms (e.g. CRs and/or turbulence).
In Figure~\ref{CO76-CO10} we add a dashed black vertical line splitting the values of the $\rm \frac{CO(7-6)}{CO(1-0)}$ ratio  between galaxies where other evidence demonstrated photon-heated molecular gas (low values), and  objects  with large ratios like the GC, the NGC253 nucleus and NGC 6240, systems where extensive studies of well-sampled molecular SLEDs concluded non-FUV photon-driven power sources for their molecular gas reservoirs.

We detect a slight trend with the IR luminosity, i.e. lower $\rm \frac{CO(7-6)}{CO(1-0)}$ are associated with galaxies with lower $\rm L(IR)$: values between 7 and 20 for luminous infrared galaxies (LIRGs) with $\rm L(IR)>11.5 ~L_\odot$, values between 2 and 7
for IRGs with $\rm <11.0 < L(IR) <11.5~ L_\odot$, and lower values  for galaxies with $\rm L(IR)<11.0~ L_\odot$. The targeted lenses are characterised by large values of the $\rm \frac{CO(7-6)}{CO(1-0)}$ ratio and have intrinsic
IR luminosity, $\rm L(IR) \geq 10^{13} L_\odot$. G12v2.43 has a $\rm \frac{CO(7-6)}{CO(1-0)}$ value similar to the local starbursts and the Galactic centre, while the same ratio for the two other lenses is extreme and shows even more extreme conditions than ultraluminous IR galaxies (ULIRGs).
We note that the value shown in Figure~\ref{CO76-CO10} for NBv1.78 is actually $\rm \frac{CO(7-6)}{CO(3-2)}$, still sensitive to
total (warm, dense)/total $\rm H_2$ gas mass, but  the corresponding actual $\rm \frac{CO(7-6)}{CO(1-0)}$ can
 be lower by a factor of up to $\sim 3$, depending of the global CO(3--2)/(1--0) line ratio ($\sim $0.3 for cold non-star-forming gas).

Low  values of the CO ratios are typical of low-infrared luminosity star-forming galaxies whose CO SLEDs are consistent with excitation by either photon-dissociation regions (PDRs) or mechanical excitation processes such as shocks and turbulence \citep[and references therein]{pap+12a,kam+17}.
In more luminous galaxies, LIRGs and brighter, this ratio and the entire CO SLED, often cannot be
explained solely by PDRs because the ratio of the brightness of the high-J CO lines to lower-J lines would be too low without extreme densities, $\rm n>10^5 cm ^{-3}$, and the far-infrared emission would be too faint given the CO line luminosities.  \citet{kam+16} report that the average SLEDs show increasing mid- to high-J CO luminosity relative to CO J = 1--0, from a few to $\sim$100,
with increasing L(IR). Even for the most luminous local galaxies, the high-J to J = 1--0 ratios do not exceed 180.
In the Galaxy such  CO line ratios can be that high only near compact hot regions in Galactic Molecular Clouds (GMCs),  results of strong and  localised excitation by intense FUV radiation from OB stars \citep{pap+12a}.

In high-luminosity galaxies, such as those in this work, different sources of excitation must be considered. One of these is the presence of the X-ray dominated regions (XDRs) where  the chemistry is driven, not by  FUV photons, but by X-ray photons that are able
to penetrate deeper into the cloud without efficiently heating the dust at the same time. These X-rays are mostly produced by
AGNs or in areas of extreme massive star formation. CRs can also heat the gas by penetrating into cloud centres,
similarly to X-rays, and are typically produced by supernovae \citep{bis+15}. Mechanical heating is another efficient source of gas heating.
This is commonly attributed to turbulence in the ISM, which may be driven by supernovae, strong
stellar winds, jets, galaxy mergers, cloud–cloud shocks, shear in the gaseous disc, or outflows \citep{mei+13,ros+15}.
These mechanisms, like main heating agents acting volumetrically on molecular gas clouds, unlike FUV photons,  can
create large mass fractions of $\rm M_{WD}/M_{tot}(H_2)$. Such large  masses of dense and warm  gas are typical
 in merger-driven local starbursts \citep{pap+12a}, i.e. only in extreme environments where this gas fraction can reach roughly ten times the value found in quiescient galaxies \citep{pap+12a,pap+14,lu+17}.

Two of the three targets also show  very large linewidths, so at least for G12v2.30 and NBv1.78 it can be interpreted as a possible sign of merger-driven velocity fields with expected large mechanical energy available as a source of energy for their molecular gas. The presence of an AGN seems to be excluded as
discussed in \citet{omo+13} and \citet{yan+16} 
from the measured 1.4 GHz radio fluxes. However, recently an AGN signature has been found in one of the images of a lensed object, SDP9, \citep{mas+17} and a hidden AGN contribution cannot be fully excluded.
In the Local Universe, high-excitation galaxies with large linewidths are not necessarily associated with high AGN contributions and galaxies with large AGN contributions do not necessarily display  large linewidths \citep{ros+15}, while low-excitation galaxies have smaller linewidths;  this narrowness is interpreted as the sign of radiative energy as the major source of excitation.

Our three targets have gravitational lens magnifications $\mu$ of about 10 \citep{bus+13,har+12}, determined from modelling the SMA continuum data at 880$\mu m$.  From these models alone, we cannot estimate how much the differential lensing affects this result. If CO(7--6) and CO(1--0) emissions come from different and decoupled regions (not overlapping along the line of sight),  the strong lensing effect may act differently on the two emissions, the resulting magnification would be different, and the ratio would be altered.
\citet{yan+17} suggest that high-J CO lines may be magnified by a factor of at least 1.3 above the overall lensing magnification for CO(1--0), whose emission is expected to be more extended. The magnification factor by \citet{yan+17} is derived by comparing the CO(1–0)/CO(3–2) ratio between a sample of unlensed SMGs \citep{bot+13} and the H-ATLAS lensed SMGs. We  expect that high-J CO would be more compact, leaving this ratio  a lower limit, as stated above.
Even if we allow the ratios shown in Figure~\ref{CO76-CO10} to decrease by this factor, the large detected excitation cannot be explained by differential lensing alone. It is worth noticing that if this were the case, G12v.2.43 would fall within the boundaries of a photon-dominated ISM.

\begin{figure}[h]
   \centering{}
   \resizebox{\hsize}{!}{\includegraphics{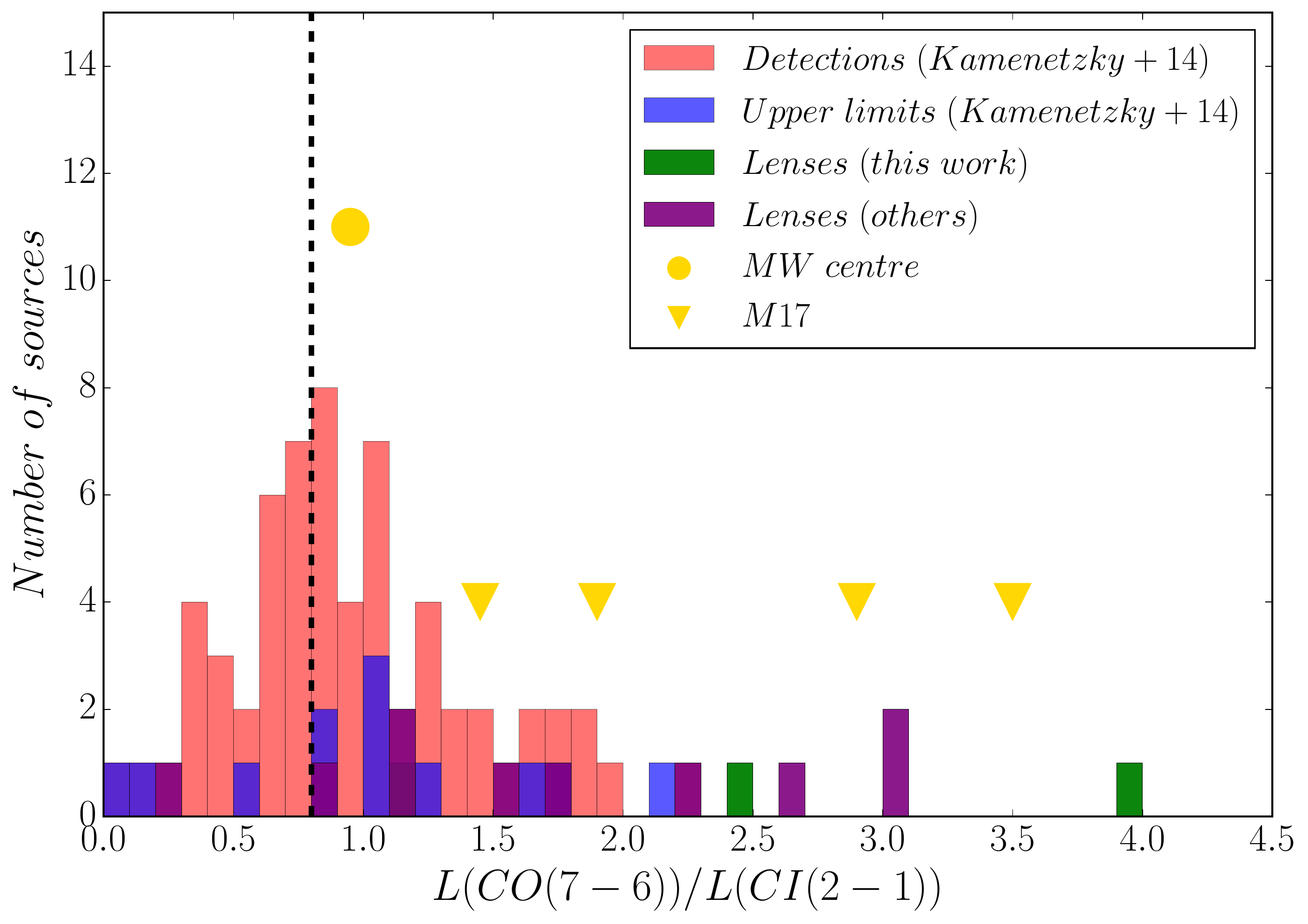}}
   \caption{Number of sources with a given CO(7--6)/CI(2--1)  luminosity ratio (in units of K km~s$^{-1}$ pc$^2$), colour- coded as in Figure~\ref{CO76-CO10}. Data for local galaxies are taken from \citet{kam+14}. A slight trend with IR luminosity is detected for the
CO(7--6)/CI(2--1) ratio: higher values correspond to larger $\rm L(IR)$. For comparison with values found in star-forming regions of the Galaxy, the range of ratios of M17 are also shown \citep{per+10}.
The blue bars correspond to upper limits (the objects have an upper limit on the CO(7--6) line emission), while for the two lenses G12v2.43 ($\sim$1.1) and NBv1.78 ($\sim$4) the green bars correspond to lower limits and the corresponding ratio is higher. Also plotted are the values of other lensed objects reported in \citet{yan+17}. The vertical dashed line indicates the value of the ratio $\rm \frac{CO(7-6)}{CI(2-1)}$ below which the ISM is dominated by the FUV heating, while those at higher values need other excitation mechanisms to explain the high value of the $\rm \frac{CO(7-6)}{CI(2-1)}$.}
              \label{CO76-CI21}%
    \end{figure}

In Figure~\ref{CO76-CI21} we report the values of the ${\rm \frac{CO(7-6)}{CI(2-1)}}$ of the lensed galaxies compared with values for the Local Universe \citep{kam+14,ros+15}. 
Also shown  in the figure are the same ratios taken from the observations of similar objects by \citet{wal+11,yan+17}.
\citet{pap+12b} suggested that the ratio between the high-J CO line and the \CI~can be used as evidence for the star-forming mode, indicating whether a system is merger driven (large values) or disc-like (low values), where the difference can be up to a factor of 10 between these two types of systems. 
\CI(2--1) is a good $\rm H_2$ mass tracer (as long as $T_k \geq 30 K$), even if \CI(2--1) (or \CI(1--0)) has not yet been well calibrated as $\rm H_2$ global mass tracers \citep{joa+17}.

For  disc-dominated environments the ${\rm \frac{CO(7-6)}{CI(2-1)}}$ ratio is expected to be lower than or roughly one, with the lowest value in the Outer Galaxy and
quiescient clouds, while in merger-driven starbursts (ULIRGs and in the Galactic centre) this ratio may be a factor of ten larger.
\citet{pap+12b} suggest using this ratio as well as a tracer of the ratio between the molecular mass in dense environments and the total molecular mass, $\rm M_{WD}/M_{tot}(H_2)$ \citep[][and references therein]{pap+12b}.
However, any differential lensing effects between the CI, which is easier to be excited, and the more extended emission with respect to the CO(7--6) emission would lower this ratio. If we were to consider the same factor as assumed for the CO(1--0) line, 1.3 less magnification than the higher-J CO lines for the \CI(2--1), the corresponding ratio ${\rm \frac{CO(7-6)}{CI(2-1)}}$ would be lower.

Even keeping in mind this effect, we can claim that most of the  lensed high-redshift objects lie in the range characteristic of large molecular gas excitation.
Neutral atomic carbon remains abundant in H$_2$ gas over a large  gas density range $100\leq n\leq 10^4 ~cm^{-3}$, i.e.
the bulk of mass of typical molecular clouds. It only becomes markedly less abundant at higher densities
where C becomes increasingly locked in CO \citep{glo+15}, but also where much less H$_2$ gas mass resides
in typical GMCs. However, for  enhanced CR energy densities in these high-redshift galaxies C can remain abundant even at higher gas densities \citep{bis+15}.  

\subsection{Gas masses}

We provide a rough estimate of $\rm H_2$ mass using the expression in \citet{pap+05} \citep[see also][]{wei+05,ala+13,bot+17,pop+17}. However, a great source of uncertainty stems from the assumed excitation conditions that determine the gas excitation function, $Q(n,T)$, which cannot be estimated without knowing  the \CI(1--0) luminosity. 
Although the \CI(1--0) is a better tracer of the molecular mass because of the smaller deviation of the $Q(n,T)$ value with respect to the local thermodynamical equilibrium (LTE) conditions 
( $\frac{Q_{1-0}}{Q_{1-0}^{(LTE)}}\sim 0.35-1$ $\frac{Q_{2-1}}{Q_{2-1}^{(LTE)}}\sim 0.15-1$ , \citet{pap+04b}), here we make  the following assumptions: (1) typical density of $100<n<5~10^4 ~cm^{-3}$, (2)  kinetic temperature $T_k=20-60K$, and (3) $T_k$ equivalent to the dust temperature. The third has been estimated from the dust spectral energy distribution and the far-infrared (FIR) luminosity \citep{zha+18} and the results are given in Table~\ref{table:results}  where  the gas mass values computed for the three targets used in this work are also listed. 

\begin{table*}
\label{mass}
\caption{Magnification uncorrected values of the neutral carbon and molecular masses}             
\label{table:results}      
\centering          
\begin{tabular}{c c c c c}     % 5 columns 
\hline\hline       
\\
Galaxy & $\rm T_{dust}$ \tablefootmark{a} & $\rm M(CI)$ & $\rm M(H_2)$%\tablefootmark{b}
   & $\rm M(H_2)$ \tablefootmark{b}\\ 
            & ($\rm K$) & ($10^7 \rm M_\odot$) &($10^{11} \rm M_\odot$) &($10^{11} \rm M_\odot$) \\

\hline    
\\
 G12v2.30 & 34 & 17$\pm$4 & 9.4$\pm$2.3 & 7.5$\pm$0.5\\  
 G12v2.43 & 31 & $<$5.4 & $<$3.0 & 1.3$\pm$0.8 \\  
 NBv1.78 & 50 & 9.2$\pm$6.5 & 5.1$\pm$3.6 & $4.6\pm0.5$ \\  % 
\\
\hline           
\end{tabular}
\tablefoot{\\
\tablefoottext{a}{data from \citet{zha+18};}
%\tablefoottext{b}{}
\tablefoottext{b}{data from CO(1--0) observations in \citet{har+12}. Gas masses in Col. 4 are derived from CI mass as in the text, and in Col. 5 from CO(1--0) luminosity. The
 conversion factor between $\rm L(CO)^\prime$ and $\rm M(H_2)$ used is $\rm 0.8 M_\odot (K~km~s^{-1}~pc^2)^{-1}$. For NBv1.78 estimated from CO(3--2) from \citet{yan+17}. Values are not corrected for magnification (see Table~\ref{table:data}).}

}
\end{table*}

The $\rm H_2$ mass found assuming a carbon abundance (in mass, relative to the molecular hydrogen) of ${\rm X[CI] = M[CI] = 6 M(H2) = 3~10^{-5}}$ \citep[][and references therein]{pap+04a,pap+04b,wei+05,bot+17,pop+17} is compared to the molecular mass estimated from the CO line luminosity \citep{har+12,yan+17,zha+18}.

Values in Table~\ref{mass} are not corrected for magnification because we are not able to define it for \CI. Our goal here is to compare the values found with the different tracers to show that \CI~can be used to estimate the molecular mass. 
With all the uncertainties listed above, the agreement between the two estimations of the molecular masses is quite satisfactory.

\subsection{Line to infrared luminosity ratios}

\begin{figure}[h]
   \centering{}
   \resizebox{\hsize}{!}{\includegraphics{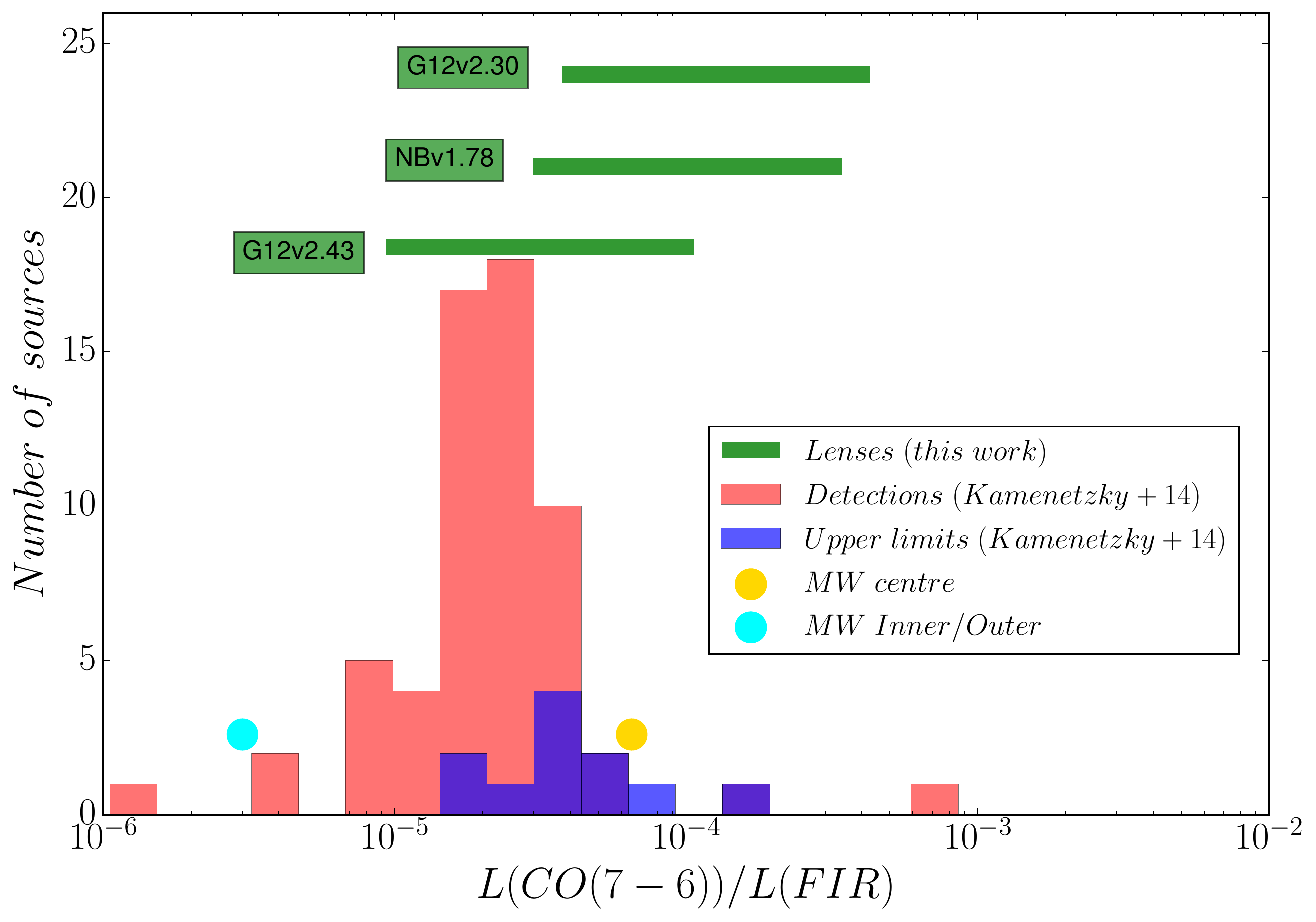}}
   \caption{Number of sources with a given CO(7--6)/FIR luminosity ratio, colour-coded as in  Figure~\ref{CO76-CO10}. %~\ref{CO76-CO10}
Data are taken from \citet{kam+14}. Values for the lens ratios show a range containing the observed values \citep{zha+18} and an uncertainty of a factor of 10, likely comprehensive of the unknown effect of the gravitational lensing effect. 
For comparison, the value of the Galaxy centre, and inner and outer Galaxy are also shown  \citep{fix+99}. }
              \label{CO76-FIR}%
    \end{figure}

In Figure~\ref{CO76-FIR} we show the distribution of the values of the luminosity ratio ${\rm CO(7-6)/FIR}$ for the lensed and the local galaxies \citep{kam+16}. This ratio spans a wide range of values from $10^{-6}$~to~$10^{-3}$, with a median of $3.5~10^{-5}$, but most of the galaxies, including the ULIRGs, have values between $10^{-5}$ and $10^{-4}$. The lowest value bins are occupied by galaxies such as NGC1097, CenA, NGC891, while the largest value is that of a dwarf low-metallicity galaxy, NGC 4560.

If we very conservatively allow a factor of ten uncertainty because of the unknown effect of the gravitational lensing on the
CO(7--6) emitting region and the overall FIR luminosity (which is likely due to the emission of the interstellar dust), the observed values for the three lenses show a wide range.
The two lenses NBv1.78 and G12v2.30 have values largely exceeding the range defined by the local galaxies, consistent with that observed in NGC 6240 \citep{mei+13} and the Galaxy centre but still lower than those of NGC 4569.
G12v2.43 has a value of $(0.1-1.3) \times 10^{-4}$ consistent with the characteristic values  of the local IR galaxies,  a result which confirms previous finding for SMGs and local galaxies \citep{lu+15}.
The high values of ${\rm CO(7-6)/FIR}$ (and also of ${\rm CI(2-1)/FIR}$) in NBv1.78 and G12v2.30 are consistent with a chemistry driven by shocks \citep{per+10}. We note that shock, CR, or turbulent heating, quite unlike heating by FUV photons, do heat the dust as effectively as they do
 gas. \citet{mei+13} find similar values in NGC6240, which they interpret as being related to shocks that compress the gas and heat it to higher temperatures, while not affecting the energy budget for the  dust reservoir that remains  much cooler.
An even higher value is seen in NGC 4569,  a low-IR luminosity galaxy in the Virgo cluster, which shows evidence of ram pressure stripping, with a deficiency of atomic hydrogen but with a large presence of molecular gas ($5~ 10^9 \rm M_\odot$) \citep{bos+16}.
\par\noindent
The ${\rm CO(7-6)/FIR}$ luminosity ratio shows a slight dependence on the IR luminosity as seen in other SMGs and lensed objects \citep{wal+11,lu+15,yan+17}, but overall it is very cumbersome to quantify this dependence in cases where the presence of an AGN altering either the FIR or both luminosities cannot be excluded.

The $\rm CO(7-6)$ luminosity is too high to be associated only with star formation rate processes as seen in local ULIRGs \citep{lu+15}, but as discussed above without a detailed lensing model it is impossible to quantify the boosting of the different emitting components.


\section{Conclusions}

Three strongly lensed  galaxies at redshift $z\sim3$ have been observed and detected with APEX/SEPIA5. The CO(7--6) emission
has been detected in all three objects, while CI(2--1) in G12v2.30, marginally in NBv1.78 and not detected in G12v2.43. 

The observed global  $\rm CO(7-6)/CI(2-1)$, $\rm CO(7-6)/CO(1-0),$ and $\rm CO(7-6)/L(FIR)$ luminosity ratios, when compared with well-studied local galaxies, show evidence for some mechanism other than the FUV heating by star formation as the dominant power source for their molecular ISM. 
Mechanical energy in the form of shocks and/or strong turbulence and/or high CR energy densities 
can maintain large amounts of very warm and dense gas, which would imply that
the average initial conditions of star formation may no longer be those found in less vigorously star-forming galaxies.

We have computed the molecular masses from the CI luminosity and find it in agreement with that derived from $\rm CO(1-0)$
within the uncertainties of both methods.

Higher angular resolution imaging together with a detailed lensing model is required to examine in detail the effect of the magnification on the different gas tracers,
to discern the relative distributions of warm dense gas associated with star formation with respect to cooler lower density gas (therefore altering their instrinsic ratios),
and also to probe the dynamics of these remarkable objects.

\begin{acknowledgements}

We would like to thank both the SEPIA team and the APEX operating team on site for their hard work in making the SEPIA commissioning a success and for conducting PI observations (under PI project 098.F-9302).
We would like to thank the referee for the suggestions that improved the readability of the paper.
CY was supported by an ESO Fellowship.
APEX is a collaboration between the Max-Planck-Institut fur Radioastronomie, the European Southern Observatory (ESO), and the Onsala Space Observatory. SEPIA is a collaboration between Sweden and ESO.      

\end{acknowledgements}

% WARNING
%-------------------------------------------------------------------
% Please note that we have included the references to the file aa.dem in
% order to compile it, but we ask you to:
%
% - use BibTeX with the regular commands:
%   \bibliographystyle{aa} % style aa.bst
%   \bibliography{Yourfile} % your references Yourfile.bib
%
% - join the .bib files when you upload your source files
%-------------------------------------------------------------------

\end{document}